\documentclass[sigconf]{acmart}

\usepackage{algorithmic}
\usepackage{algorithm}
\usepackage{textcomp}
\usepackage{pifont}
\usepackage{nicefrac}
\usepackage{multirow}

\AtBeginDocument{%
  }

\definecolor{lightgreen}{RGB}{0,180,0}
\definecolor{lightblue}{RGB}{150,210,255}

\definecolor{linkblue}{RGB}{0, 80, 160}

\copyrightyear{2026}
\acmYear{2026}
\setcopyright{cc}
\setcctype{by}
\acmConference[ISLPED '26]{ACM/IEEE International Symposium on Low Power Electronics and Design}{August 05--07, 2026}{Evanston, IL, USA}
\acmBooktitle{ACM/IEEE International Symposium on Low Power Electronics and Design (ISLPED '26), August 05--07, 2026, Evanston, IL, USA}
\acmDOI{10.1145/3816440.3818553}
\acmISBN{979-8-4007-2748-1/2026/08}

\begin{document}

\title{SpikON: A Dual-Parallel and Efficient Accelerator
for Online Spiking Neural Networks Learning}

\author{Peilin Chen}
\email{peilin@virginia.edu}
\affiliation{
  \institution{University of Virginia}
  \city{Charlottesville}
  \state{VA}
  \country{USA}
}

\author{Xiaoxuan Yang}
\email{xiaoxuan@virginia.edu}
\affiliation{
  \institution{University of Virginia}
  \city{Charlottesville}
  \state{VA}
  \country{USA}
}

\begin{abstract}
Spiking neural networks~(SNNs) have emerged as a promising paradigm for energy-efficient brain-inspired computing. However, existing online unsupervised SNN learning suffers from low training accuracy and poor scalability. Although current online supervised learning algorithms perform well on large-scale datasets and networks, the non-hardware-friendly operations hinder efficient edge deployment. In this work, we propose SpikON, the first algorithm-hardware co-design framework for efficient and scalable end-to-end online supervised SNN learning. We first propose the learnable threshold through time and scaled weight centralization through time techniques to address the inefficiency of traditional algorithms. Moreover, to reduce latency and energy consumption, we introduce the novel training dataflow and cascade computation reuse scheme for SNNs that allows concurrent forward-backward computation and temporal reuse across timesteps. We further design the dedicated SNN accelerator with a dual-parallel engine and customized SIMD-based SNN core for efficient end-to-end online learning. Experiments show that the SpikON algorithm achieves 32.2\% and 35.0\% reductions in training latency and energy consumption over the baseline, without sacrificing accuracy. Moreover, the SpikON co-design achieves 7.2x~(11.5x) and 26.8x~(15.8x) training throughput~(energy efficiency) compared with the edge Apple M4 GPU and TPU-like accelerator, respectively. The code is available at \href{https://github.com/peilin-chen/SpikON}{\textcolor{linkblue}{GitHub}}.
\end{abstract}




\maketitle

\section{Introduction}

Spiking neural networks~(SNNs) emerge as a bio-inspired computing paradigm, bridging the gap between artificial neural networks~(ANNs) and human brains~\cite{eshraghian2023training}. SNNs mimic the dynamics of biological neurons and utilize binary spikes to communicate between layers~\cite{jiang2024ndot, yin2024loas}. Due to the inherent sparsity, SNNs are regarded as more energy-efficient than ANNs and are widely used in low-power edge computing~\cite{hasssan2024spiking, kim2021revisiting, mao2024stellar}. Meanwhile, backpropagation through time~(BPTT) with surrogate gradient enables the training of deep SNNs that achieve accuracy comparable to ANN counterparts~\cite{xiao2022online, zheng2021going}.

However, BPTT suffers from significant memory cost during training, since it requires storing intermediate data at all timesteps~\cite{jiang2024ndot, meng2023towards, yin2023accurate}. BPTT is also inconsistent with online learning, where gradients are computed immediately at each timestep~\cite{xiao2022online, jiang2024ndot, meng2023towards, davies2018loihi}. Previous work modifies BPTT to enable forward-in-time SNN learning by tracking presynaptic activities~\cite{xiao2022online, jiang2024ndot} or ignoring unimportant temporal gradients~\cite{meng2023towards}. Despite their success on large-scale datasets, we identify three main challenges in deploying scalable online supervised SNN learning algorithms on edge devices. 

\begin{figure}[tb]
    \centering
    \setlength{\abovecaptionskip}{0pt}
    \includegraphics[width=\linewidth]{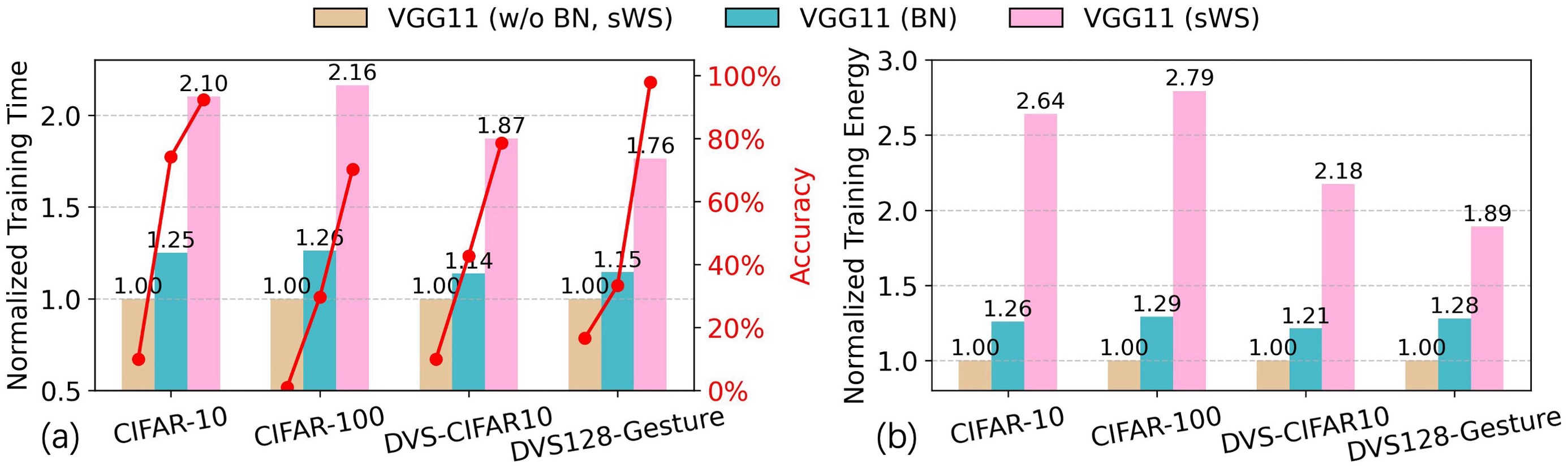}
    \vspace{-10pt}
    \caption{Online SNN training of VGG11 based on \cite{meng2023towards} under Nvidia A40 GPU. (a) Normalized training time and corresponding accuracy. (b) Normalized training energy. Timesteps: 6, 6, 10, and 20 for the datasets in the figure.}
    \Description{fig1}
    \label{fig1}
    \vspace{-12pt}
\end{figure}

\textit{\underline{(1) Non-hardware-friendly normalization.}} In online SNN learning scenarios, training samples are provided to the network sequentially~(one single sample at a time)~\cite{xiao2022online, davies2018loihi, lobo2020spiking}. However, batch normalization~(BN)~\cite{ioffe2015batch} becomes ineffective with batch size ${=}$ 1, as it relies on large batches to estimate the dataset statistics~\cite{qiao2019micro}. To solve this issue, prior work proposes the scaled weight standardization~(sWS) to replace BN~\cite{qiao2019micro, xiao2022online, jiang2024ndot, meng2023towards}. Although online SNN training with sWS can achieve comparable accuracy with BPTT, it introduces $2.0\times$ and $2.4\times$ higher training time and energy on average~(Fig.~\ref{fig1}(a) and (b)) compared to the vanilla network~(without BN and sWS). \textit{\underline{(2) Inherent multi-timestep computation.}} Compared to ANNs that process inputs in a single shot, SNNs require computation over multiple timesteps during both forward and backward passes~\cite{chowdhury2021one}. This inherent temporal property of SNNs increases both training latency and energy consumption due to the repetitive computations across timesteps. Moreover, the overall training overhead of SNNs scales proportionally with the number of timesteps~\cite{yin2022sata}. \textit{\underline{(3) Lack of efficient hardware support.}} Most online learning systems are designed for the unsupervised spike timing-dependent plasticity~(STDP) algorithm. However, STDP suffers from low training accuracy and poor scalability for large networks and datasets~\cite{davies2018loihi, li2021fast, qiao2019neuromorphic, sun2022energy}. Although one recent work~\cite{siddique2023low} designs hardware for its proposed online supervised SNN learning algorithm, it remains constrained to small-scale datasets such as MNIST~\cite{lecun2002gradient}.

To address the aforementioned challenges, we develop an algorithm-hardware co-design framework named SpikON, featuring high-throughput and energy-efficient training. To the best of our knowledge, SpikON is the \textit{first} system that enables efficient and scalable online supervised SNN learning. Specifically, the main contributions are summarized as follows:
\begin{enumerate}
    \item We design the dedicated accelerator to support the end-to-end online SNN training, with a dual-parallel learning engine and customized SIMD-based SNN core.
    \item We propose the hardware-friendly learnable threshold through time (LTTT) and scaled weight centralization through time (sWCTT) approaches to overcome the inefficiency of sWS, without sacrificing training accuracy.
    \item We propose the bi-directional temporal parallel~(BTP) training dataflow that enables the concurrent forward and backward computation across all timesteps, effectively reducing SNN training latency.
    \item We introduce the cascade temporal computation reuse~(CTCR) scheme that utilizes the spike activation similarity between adjacent timesteps to reuse partial results, thus reducing redundant operations and training energy.
\end{enumerate}

\section{Background}

\begin{figure}[tb]
    \centering
    \setlength{\abovecaptionskip}{0pt}
    \includegraphics[width=\linewidth]{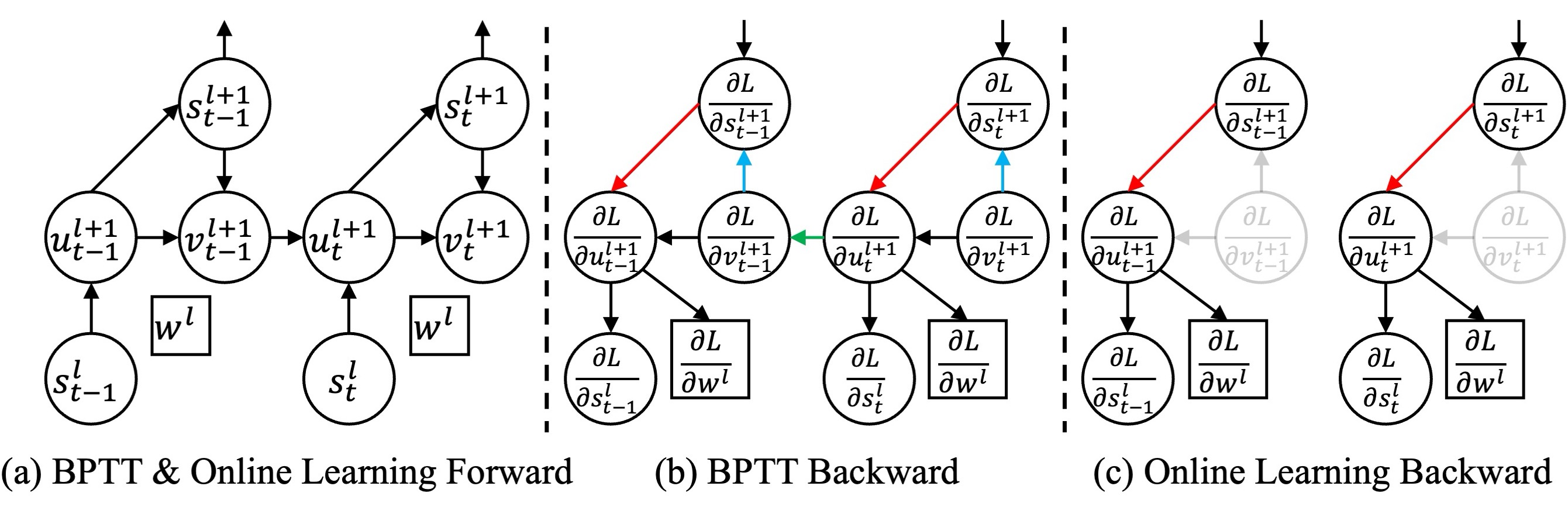}
    \vspace{-10pt}
    \caption{Comparison of the forward and backward processes between BPTT and online SNN learning. $s^l_t$, $u^l_t$, and $v^l_t$ denote the spike output, membrane potential after charging, and membrane potential after resetting of layer $l$ at timestep $t$. $L$ and $w^l$ represent the loss and weights of SNN in layer $l$.}
    \Description{fig2}
    \label{fig2}
    \vspace{-10pt}
\end{figure}

\subsection{Online Learning Algorithm for SNNs}
\label{section2_1}
BPTT and online SNN learning follow the same forward propagation flow~(Fig.~\ref{fig2}(a)), which is implemented based on leaky integrate-and-fire~(LIF) neurons~\cite{fang2023spikingjelly} modeled with three stages: charging~($u^{l+1}_t$), firing~($s^{l+1}_t$), and resetting~($v^{l+1}_t$). However, BPTT backpropagates gradients through the temporal and spatial directions, as shown in Fig.~\ref{fig2}(b). This process can be formulated as follows:
\begin{equation}
\frac{\partial L}{\partial w^l} = \sum_{t=0}^{T-1} \sum_{j \leq t} \frac{\partial L}{\partial w^l_{j}},with\text{ }w^{l}_{j}=w^{l}\ \forall j\text{},
\end{equation}
where $T$ is the timestep. When $j$ is equal to $t{-}1$, we obtain 
\begin{equation*}
\resizebox{\linewidth}{!}{$
\frac{\partial L}{\partial w^l_{t{-}1}}{=}\smash\{\frac{\partial L}{\partial s^{l+1}_{t-1}}\textcolor{red}{\frac{\partial s^{l+1}_{t-1}}{\partial u^{l+1}_{t-1}}}{+}\frac{\partial L}{\partial u^{l+1}_t} \textcolor{lightgreen}{\frac{\partial u^{l+1}_t}{\partial v^{l+1}_{t-1}}}(
\frac{\partial v^{l+1}_{t-1}}{\partial u^{l+1}_{t-1}}{+}\textcolor{lightblue}{\frac{\partial v^{l+1}_{t-1}}{\partial s^{l+1}_{t-1}}}\textcolor{red}{\frac{\partial s^{l+1}_{t-1}}{\partial u^{l+1}_{t-1}}})\smash\}\frac{\partial u^{l+1}_{t-1}}{\partial w^{l}_{t-1}},$}
\end{equation*}
where $\textcolor{lightgreen}{\nicefrac{\partial u^{l+1}_t}{\partial v^{l+1}_{t-1}}}\in(0,1)$ is the leaky constant $\beta$. Note that a surrogate function~\cite{eshraghian2023training, neftci2019surrogate} is used to approximate the non-differentiable $\textcolor{red}{\nicefrac{\partial s^{l+1}_{t-1}}{\partial u^{l+1}_{t-1}}}$. We can observe that BPTT cannot compute the gradient instantly at each timestep. To solve this issue, prior work~\cite{meng2023towards} shows that temporal gradients are unimportant during backpropagation and proposes the spatial learning through time~(SLTT) method. Therefore, weight gradients can be calculated by $\frac{\partial L}{\partial w^l}{=}\sum_{t=0}^{T-1} \frac{\partial L}{\partial s^{l+1}_{t-1}}\\\textcolor{red}{\frac{\partial s^{l+1}_{t-1}}{\partial u^{l+1}_{t-1}}}\frac{\partial u^{l+1}_{t-1}}{\partial w^{l}}$~(Fig.~\ref{fig2}(c)). In this way, gradients at each timestep are computed independently without relying on $\textcolor{lightgreen}{\nicefrac{\partial u^{l+1}_t}{\partial v^{l+1}_{t-1}}}$, which aligns with the property of online SNN learning. 

\subsection{BN and sWS in SNNs}
BN is introduced to handle the internal covariate shift issue in deep network training~\cite{ioffe2015batch}. However, BN fails to work well when applied to online SNN learning~(Fig.~\ref{fig1}(a)). To overcome this limitation, sWS~\cite{qiao2019micro, meng2023towards} is proposed to replace BN by normalizing weights rather than activations. The sWS is defined as:
\begin{equation}
\hat{w}_{i,j} = \gamma \cdot 
\frac{w_{i,j} - \mu_{w_{i, .}}}
{\sigma_{w_{i, .}} \sqrt{N}}, \label{equation2}
\end{equation}
where $w_{i,j}$, $\hat{w}_{i,j}$, $N$, and $\gamma$ denote the original, standardized weights, fan-in size, and fixed scale value, respectively. The mean~($\mu_{w_{i, .}}$) and variance~($\sigma_{w_{i, .}}$) are computed along the input channel dimension. Although previous work~(e.g., SLTT) demonstrates that sWS can improve training accuracy under batch size ${=}1$, it introduces substantial latency and energy overheads~(Fig.~\ref{fig1}). This motivates us to design a more efficient algorithm tailored for online SNN learning.

\section{SpikON Algorithm Design}
\label{section3}
\begin{figure}[tb]
    \centering
    \setlength{\abovecaptionskip}{0pt}
    \includegraphics[width=\linewidth]{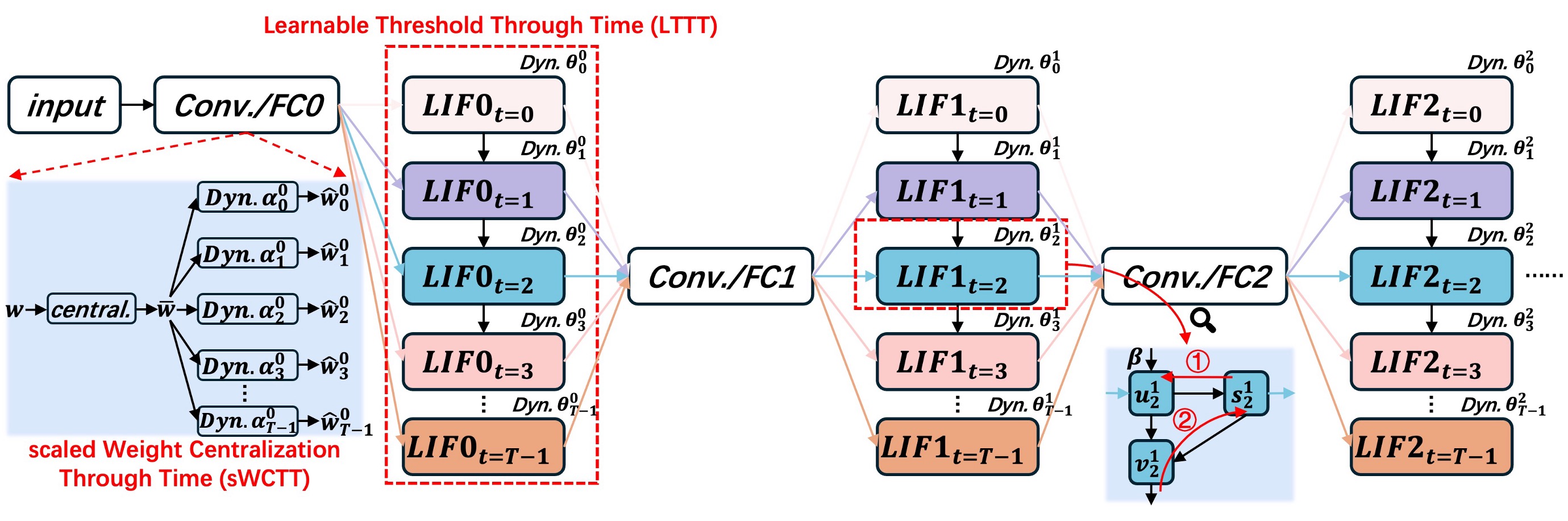}
    \vspace{-10pt}
    \caption{Overview of the proposed learnable threshold through time~(LTTT) and scaled weight centralization through time~(sWCTT) methods. Conv.: convolutional layer. FC: fully-connected layer. Dyn.: dynamic.}
    \Description{fig3}
    \label{fig3}
    \vspace{-16pt}
\end{figure}

Prior online SNN training algorithms~(e.g., SLTT) directly adopt sWS, which was originally designed for deep ANNs, to replace BN. However, sWS overlooks the unique LIF neuronal dynamics and temporal dependencies of SNNs. This motivates us to rethink normalization for SNNs: \textit{can we leverage the intrinsic characteristics of SNNs to achieve normalization-free and efficient online learning?} To this end, we propose two complementary approaches on SNN activations and weights, namely LTTT and sWCTT.

\subsection{Learnable Threshold Through Time}
Unlike previous SNN works~\cite{tan2021improved, rathi2021diet, hasssan2024spiking, wang2022ltmd} that use a fixed or dynamic threshold parameter~($\theta$) shared across all timesteps, LTTT assigns each timestep in the LIF neuron an independent learnable $\theta$~(Fig.~\ref{fig3}). This design approximates activation normalization~(e.g., BN) by controlling spike firing and balancing membrane potential through time. 

As illustrated in Fig.~\ref{fig3}, a set of $T$ learnable thresholds~($\theta$ for $t{=}0,\dots,T{-}1$) is introduced in the LIF neuron. The forward computation can be formulated as follows: 
\begin{align}
\textit{firing: }s_{t}^{l+1} =H(u_t^{l+1}{-}\theta_t^{l+1})&=
\begin{cases}
1,&\text{if } u_t^{l+1} \geq \theta_t^{l+1} \\
0,&\text{otherwise},
\end{cases} \label{equation3} \\
\textit{charging: }u_t^{l+1} &= \beta v_{t-1}^{l+1} + w^{l} s_t^l, \label{equation4} \\
\textit{resetting: }v_t^{l+1} &= u_t^{l+1} - s_{t}^{l+1}\theta_t^{l+1} \label{equation5} ,
\end{align}
where $\theta_t^{l+1}$ is the learnable threshold of layer $l{+}1$ at timestep $t$, and $H$ denotes the Heaviside step function. Each timestep in LTTT utilizes its own $\theta_t^{l+1}$, enabling temporal adaptation during training. The threshold updates can be calculated as:
\begin{align}
\nabla\theta_{t}^{l+1} &{=} \frac{\partial L}{\partial s_t^{l+1}}\frac{\partial s_t^{l+1}}{\partial \theta_{t}^{l+1}}{=}{-}\frac{\partial L}{\partial s_t^{l+1}}H^{'}(u_t^{l+1}{-}\theta_t^{l+1}), \label{equation6}\\
\theta_t^{l+1} &= \theta_t^{l+1} {-} \eta\nabla\theta_{t}^{l+1},
\end{align}
where $\eta$ represents the learning rate. Due to the non-differentiability of $H$, we adopt triangle surrogate function to approximate $H^{'}(u_t^{l+1}{-}\theta_t^{l+1})$. Moreover, equation~(\ref{equation6}) focuses on the backward pass \textcolor{red}{\textcircled{1}} in Fig.~\ref{fig3} because the temporal gradient is ignored in online learning. Note that the overhead introduced by LTTT is minimal, as it only adds $T$ trainable parameters per LIF neuron, and the required $\nicefrac{\partial L}{\partial s_t^{l+1}}$ can be directly reused from the weight update process. Therefore, both the parameter and computation costs are negligible. 

\begin{algorithm}[tb]
    \caption{One training iteration of the SpikON algorithm}
    \label{algorithm1}
    \renewcommand{\algorithmicrequire}{\textbf{Input:}}
    \renewcommand{\algorithmicensure}{\textbf{Output:}}
    
    \begin{algorithmic}[1]
        \REQUIRE Network parameters $w^l$, $\theta^l_t$, $\alpha^l_t$; Total layer number $N$; SNN timestep $T$; Learning rate $\eta$; Other required training data.
        \STATE \textbf{Initialize:} $\nabla w^l$=0. %$\nabla \theta^l$=0, and $\nabla \alpha^l$=0.
        %\ENSURE Trained SNN parameters $w^l$, $\theta^l_t$, $\alpha^l_t$.
        \FOR{$t=0,1,\dots, T{-}1$}
            \FOR{$l=0,1,\dots,N{-}1$} 
                \STATE \textit{Forward pass:} Calculate $u^l_t$, $v^l_t$, and $s^l_t$ using equation (\ref{equation4}), (\ref{equation5}), (\ref{equation3}) (weights are rescaled by equation (\ref{equation8}));
            \ENDFOR
            \FOR{$l=N{-}1,\dots,1, 0$}
                \STATE \textit{Backward pass:} Calculate the instantaneous loss $L$;
                \STATE Calculate instantaneous gradients $\nabla w^l_t$, $\nabla \theta^l_t$, and $\nabla \alpha^l_t$ by equation (\ref{equation9}), (\ref{equation6}), and (\ref{equation10});
                \STATE \textit{Weight gradient accumulation:} $\nabla w^l$=$\nabla w^l$ + $\nabla w^l_t$; %$\nabla \theta^l$ = $\nabla \theta^l$ + $\nabla \theta^l_t$, $\nabla \alpha^l$ = $\nabla \alpha^l$ + $\nabla \alpha^l_t$;
            \ENDFOR
        \ENDFOR
        \STATE \textit{Parameter update:} $w^l$ = $w^l$ $-$ $\eta$$\nabla w^l$, $\theta^l_t$ = $\theta^l_t$ $-$ $\eta$$\nabla \theta^l_t$, $\alpha^l_t$ = $\alpha^l_t$ $-$ $\eta$$\nabla \alpha^l_t$;
         \ENSURE Trained SNN parameters $w^l$, $\theta^l_t$, $\alpha^l_t$.
    \end{algorithmic}
\end{algorithm}

\subsection{Scaled Weight Centralization Through Time}
Similar to how LTTT approximates activation normalization, sWCTT aims to achieve the effect of weight normalization. Specifically, the weights are first centralized by subtracting their mean value to make the distribution zero-centered. Moreover, sWCTT introduces a learnable scaling factor $\alpha_t^l$ for each timestep to rescale the centralized weights $\overline{w}^l$~(Fig.~\ref{fig3}). This mechanism regulates the weight magnitude over time and stabilizes the weight distribution during training. The sWCTT can be formulated as follows: 
\begin{align}
\textit{forward: }\overline{w}^l_{t,i,j}&{=}w^l_{t,i,j}{-}\mu_{w^l_{t,i,.}},\hat{w}^l_t{=}\alpha^l_t\overline{w}^l_{t},w^{l}_{t}{=}w^{l}\ \forall t \label{equation8}\\
\textit{backward: }\frac{\partial L}{\partial w^l_{t,i,j}}&{=}\alpha^l_t(g^l_{t,i,j}{-}\mu_{g^l_{t,i,.}}), \label{equation9} \\
\frac{\partial L}{\partial \alpha^l_{t}}&{=}\sum_{i,j}g^l_{t,i,j}\overline{w}^l_{t,i,j},\textit{ with } g^l_{t}{=}\frac{\partial L}{\partial \hat{w}^l_t}, \label{equation10}
\end{align}
where $w^l_t$, $\overline{w}^l_{t}$, and $\hat{w}^l_t$ denote the original, centralized, and scaled weights, respectively, and $g^l_t$ is the scaled weight gradient. The $i$ and $j$ represent the output and input neuron indices, respectively. Different from sWS that applies a fixed pre-computed $\gamma$~(equation~(\ref{equation2})), sWCTT assigns an independent learnable $\alpha^l_t$ to each timestep, allowing more flexible weight adjustment over time. Moreover, sWCTT eliminates the need to calculate variance and simplifies the weight gradient computation, making it more hardware-friendly for efficient online SNN learning.

\subsection{Overall Pseudocode of SpikON Algorithm}
Algorithm~\ref{algorithm1} outlines the overall training process of SpikON, which combines the proposed LTTT and sWCTT for efficient online SNN learning. Specifically, during the forward pass, the timestep-dependent $\theta^l_t$ controls the update of potentials and spike states~(Line 4). Once the forward propagation at a given timestep is completed, the backward pass computes the instantaneous loss~($L$) and corresponding gradients $\nabla w^l_t$, $\nabla \theta^l_t$, and $\nabla \alpha^l_t$~(Line 7, 8). The gradients obtained from all timesteps are subsequently accumulated and applied to update the network parameters~(Line 9, 12).

\begin{figure}[tb]
    \centering
    \setlength{\abovecaptionskip}{0pt}
    \includegraphics[width=\linewidth]{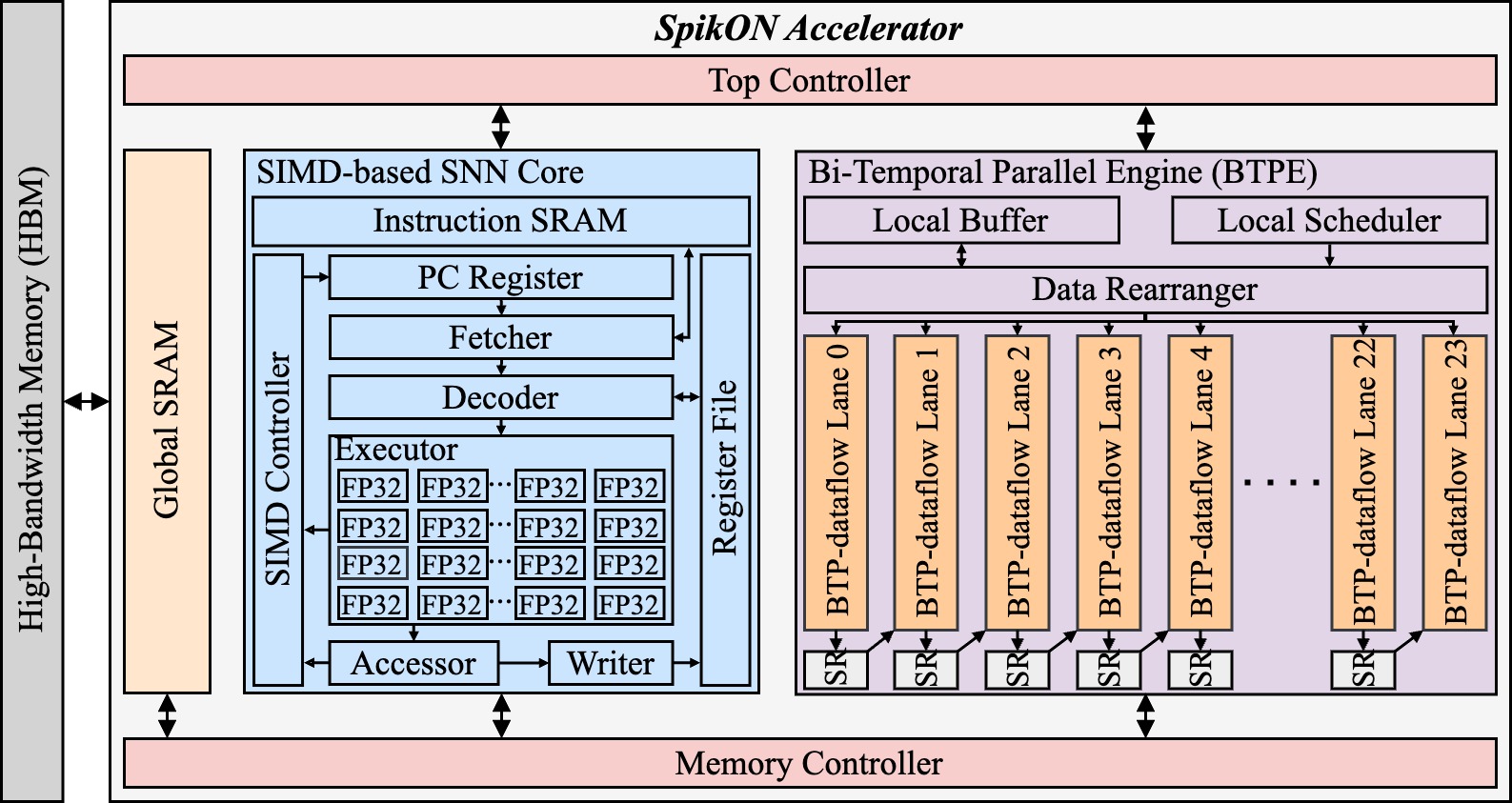}
    \vspace{-6pt}
    \caption{The overall architecture of SpikON training accelerator. The global and instruction SRAMs are 1 MB and 4 KB, respectively. Each lane is attached to a 256 KB local SRAM.}
    \Description{fig4}
    \label{fig4}
    \vspace{-14pt}
\end{figure}
\section{SpikON Hardware Architecture}

\begin{figure*}[tb]
    \centering
    \setlength{\abovecaptionskip}{0pt}
    \includegraphics[width=0.99\linewidth]{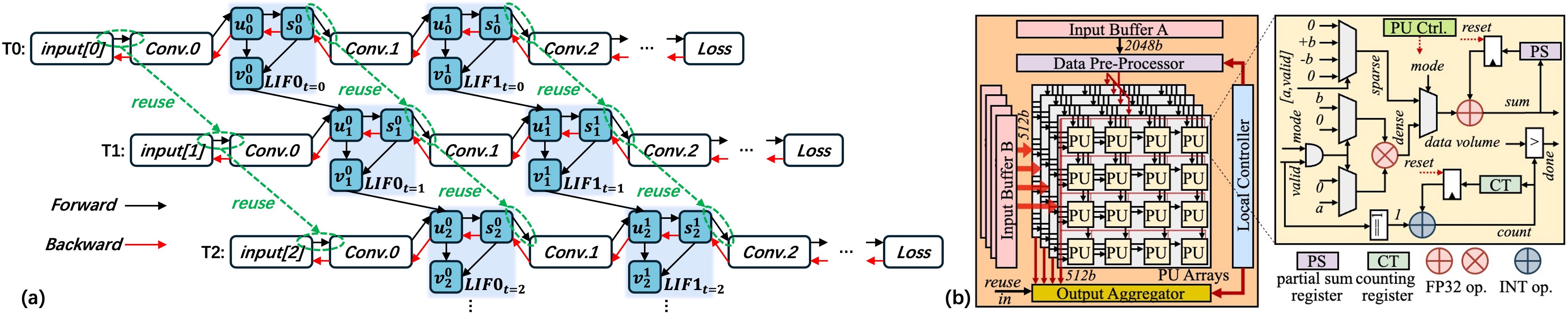}
    \vspace{-4pt}
    \caption{(a) Illustration of the proposed bi-directional temporal parallel training dataflow and cascade temporal computation reuse methods. (b) The microarchitecture of the BTP-dataflow lane. FP32: floating-point 32-bit. INT: integer.}
    \Description{fig5}
    \label{fig5}
    \vspace{-12pt}
\end{figure*}

\subsection{Overview of SpikON Architecture}
The proposed SpikON accelerator is designed to support efficient and scalable online SNN learning. As shown in Fig.~\ref{fig4}, SpikON consists of five components: bi-temporal parallel engine~(BTPE), customized SIMD-based SNN core, global SRAM, memory controller, and top controller. The BTPE performs the vector-matrix multiplications~(VMM) in both forward and backward passes, including $\nabla w$ and error signal computations. It supports bi-directional temporal parallel~(BTP) dataflow and cascade temporal computation reuse~(CTCR) scheme to reduce training latency and energy. Moreover, the SIMD-based SNN core executes element-wise and vector-wise operations, such as average pooling, softmax, and LIF neuron update. The memory controller arbitrates concurrent accesses from the SNN core and BTPE to the global SRAM, while the top controller manages the overall operation by controlling start signals for their coordinated execution.

\subsection{Bi-directional Temporal Parallel Dataflow}

\textbf{BTP training dataflow.} We design the BTP dataflow to improve online SNN training throughput while effectively handling temporal dependencies. Unlike traditional training flow that processes each timestep's forward and backward pass in order, the BTP dataflow enables concurrent forward-backward computation across all timesteps~(Fig.~\ref{fig5}(a)). Specifically, in the forward pass, computations of adjacent timesteps are interleaved to exploit temporal parallelism. For example, the $Conv.0$ layer of T0 starts first, and the $Conv.0$ layer of T1 begins when T0 proceeds to $LIF0_{t=0}$ layer. Once the potential $v^0_0$ in $LIF0_{t=0}$ is updated, it can be utilized by the next timestep, allowing T0's $Conv.1$, $LIF0_{t=1}$, and T2's $Conv.0$ to execute concurrently. This staggered execution continues across timesteps, achieving high forward parallelism. Different from the forward process, the backward pass in the SpikON algorithm does not involve temporal error propagation across timesteps~(Section~\ref{section3}). Thus, all timesteps can be executed independently, enabling fully parallel backward computation. 

\begin{figure}[tb]
    \centering
    \setlength{\abovecaptionskip}{0pt}
    \includegraphics[width=0.98\linewidth]{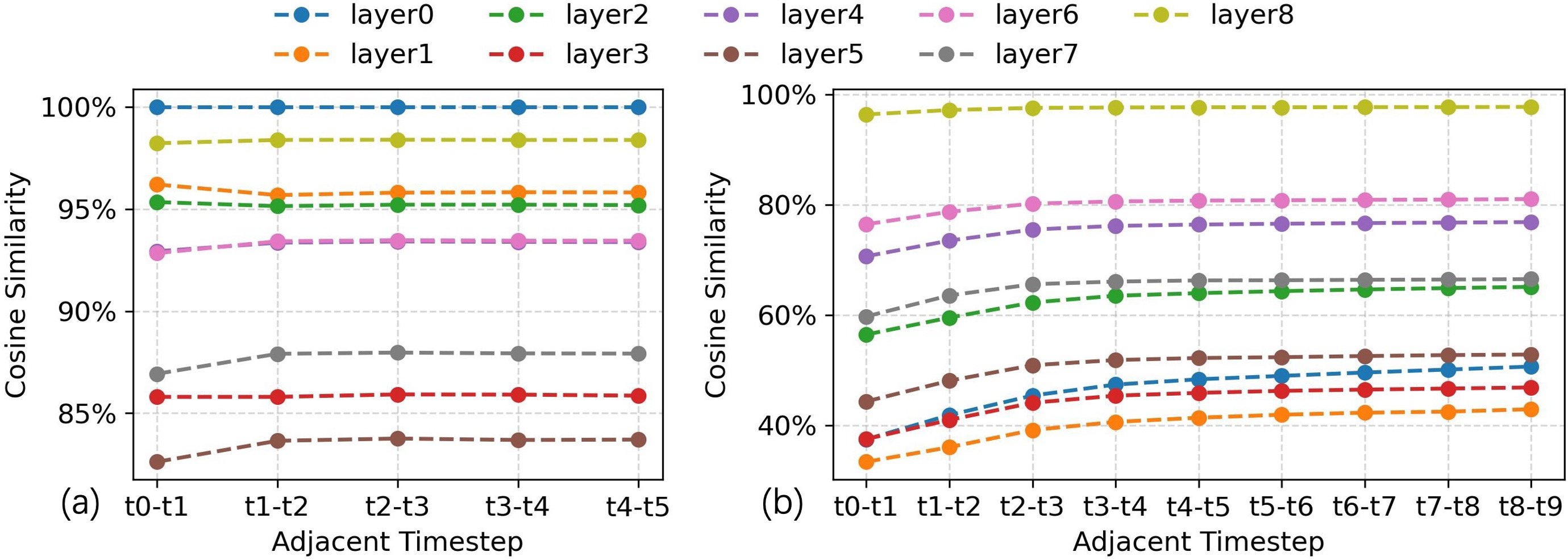}
    %\vspace{-2pt}
    \caption{Spike cosine similarity of adjacent timesteps on CIFAR-100~(a) and DVS-CIFAR10~(b) datasetes using the SpikON training algorithm~(VGG11).}
    \Description{fig6}
    \label{fig6}
    \vspace{-16pt}
\end{figure}

\textbf{Dual-parallel learning engine.} To support the proposed BTP training dataflow, we design the specialized BTPE module, as shown in the right part of Fig.~\ref{fig4}. The BTPE consists of 24 BTP-dataflow lanes, buffer, scheduler, data rearranger, and 23 local SRAMs. Each lane in BTPE is assigned to process one timestep of SNN, supporting up to 24 timesteps in parallel. When $T$ is smaller than 24, multiple lanes can be grouped to handle the same timestep. For example, for SNN with $T=6$, lanes 0, 6, 12, and 18 are used to process $T0$. The local scheduler uses the finite state machine to control data loading, distribution, result write-back, etc. Moreover, local SRAMs are utilized to enable the CTCR method by storing lane-level intermediate results for reuse~(Section~\ref{section4_3}).

\textbf{BTP-dataflow lane.} To better leverage the inherent sparsity of SNNs, the BTP-dataflow lane~(Fig.~\ref{fig5}(b)) in the BTPE is implemented to support both sparse and dense computation modes during training. Specifically, each lane consists of input buffers A and B, data pre-processor, four processing unit~(PU) arrays, output aggregator, and local controller. The PU array contains $4\times 4$ PUs that adopt an output-stationary dataflow, where each PU computes one element of the output feature map, allowing for 64-element parallel processing per lane. The output aggregator~(OA) accumulates the partial results of current PU arrays with outputs from the previous timestep. Moreover, the local controller coordinates data access to local SRAMs within BTPE and generates the $done$ signal when PU arrays and OA finish computation.

\textbf{Processing unit.} As shown in Fig.~\ref{fig5}(b), PU consists of FP32 multiplier and adder, partial sum~(PS) and counting~(CT) registers, control logic, etc. It operates in two computation modes: sparse and dense. During the sparse mode, the $mode$ signal disables the dense path while activating the sparse path. The multiplexer selects among $0$, $-b$, or $+b$ for computing potential $u$ and $\nabla w$. Note that $-b$ indicates the negative number utilized in the CTCR method. In contrast, during the dense mode, the $mode$ signal enables accumulation using results from the dense path for error signal computation. Once the processed element pairs reach the preset $data$ $volume$, the PU sets $done=1$ to indicate computation completion.

\vspace{-6pt}
\subsection{Cascade Temporal Computation Reuse}
\label{section4_3}

\begin{table*}[tb]
\centering
\caption{Performance comparison among online SNN learning methods OTTT, SLTT, and SpikON on CIFAR-10, CIFAR-100, DVS-CIFAR10, and DVS128-Gesture using NVIDIA A40 GPU.}
\vspace{-10pt}
\label{table1}
\renewcommand{\arraystretch}{0.9}
\resizebox{0.92\linewidth}{!}{
\begin{tabular}{c|ccccccc}
\toprule
\textbf{Dataset} & \textbf{Method} & \textbf{Network} & \textbf{Params} & \textbf{Timesteps} & \textbf{Accuracy} & \textbf{Training Time/Epoch (s)} & \textbf{Training Energy/Epoch (J)} \\
\midrule
\multirow{6}{*}{CIFAR-10} 
& \textbf{OTTT} & \textbf{VGG11 (sWS)} & \textbf{9.23M} & \textbf{6} & \textbf{92.47\%} & \textbf{2,829.10} & \textbf{525,395.45} \\
%& SLTT & VGG11 (w/o BN, sWS) & 9.23M & 6 & 10\% (cannot train) & 975.2 & 150,565.70 \\
& SLTT & VGG11 (BN) & 9.23M & 6 & 74.30\% & 1,219.80 & 189,402.19 \\
& \textbf{SLTT} & \textbf{VGG11 (sWS)} & \textbf{9.23M} & \textbf{6} & \textbf{92.47\%} & \textbf{2,048.03} & \textbf{397,490.18} \\
& SpikON (ours) & VGG11 (LTTT) & 9.23M & 6 & 90.88\% & 1,028.20 & 163,299.09 \\
& \textbf{SpikON (ours)} & \textbf{VGG11 (LTTT+sWCTT)} & \textbf{9.23M} & \textbf{6} & \textbf{92.24\%} & \textbf{1,395.98} & \textbf{245,739.04} \\
\midrule
\multirow{6}{*}{CIFAR-100} 
& \textbf{OTTT} & \textbf{VGG11 (sWS)} & \textbf{9.27M} & \textbf{6} & \textbf{70.16\%} & \textbf{2,886.71} & \textbf{524,567.27} \\
%& SLTT & VGG11 (w/o BN, sWS) & 9.27M & 6 & 1\% (cannot train) & 968.93 & 143,818.31 \\
& SLTT & VGG11 (BN) & 9.27M & 6 & 29.67\% & 1,223.17 & 185,793.66 \\
& \textbf{SLTT} & \textbf{VGG11 (sWS)} & \textbf{9.27M} & \textbf{6} & \textbf{70.29\%} & \textbf{2,095.40} & \textbf{401,829.07} \\
& SpikON (ours) & VGG11 (LTTT) & 9.27M & 6 & 56.68\% & 1,052.81 & 164,892.59 \\
& \textbf{SpikON (ours)} & \textbf{VGG11 (LTTT+sWCTT)} & \textbf{9.27M} & 6 & \textbf{69.84\%} & \textbf{1,387.75} & \textbf{239,152.99} \\
\midrule
\multirow{6}{*}{DVS-CIFAR10} 
& \textbf{OTTT} & \textbf{VGG11 (sWS)} & \textbf{9.23M} & \textbf{10} & \textbf{76.80\%} & \textbf{962.6} & \textbf{180,017.09} \\
%& SLTT & VGG11 (w/o BN, sWS) & 9.23M & 10 & 10\% (cannot train) & 378 & 58,993.59 \\
& SLTT & VGG11 (BN) & 9.23M & 10 & 42.80\% & 429.95 & 71,574.49 \\
& \textbf{SLTT} & \textbf{VGG11 (sWS)} & \textbf{9.23M} & \textbf{10} & \textbf{78.70\%} & \textbf{707.94} & \textbf{128,350.70} \\
& SpikON (ours) & VGG11 (LTTT) & 9.23M & 10 & 76.10\% & 387 & 61,699.13 \\
& \textbf{SpikON (ours)} & \textbf{VGG11 (LTTT+sWCTT)} & \textbf{9.23M} & \textbf{10} & \textbf{79.00\%} & \textbf{496.1} & \textbf{86,394.05} \\
\midrule
\multirow{7}{*}{DVS128-Gesture} 
& \textbf{OTTT} & \textbf{VGG11 (sWS)} & \textbf{9.23M} & \textbf{20} & \textbf{97.22\%} & \textbf{348.15} & \textbf{72,046.33} \\
%& SLTT & VGG11 (w/o BN, sWS) & 9.23M & 20 & 16.67\% (cannot train) & 118.32 & 24,906.57 \\
& SLTT & VGG11 (BN) & 9.23M & 20 & 33.33\% & 135.57 & 31,873.54 \\
& \textbf{SLTT} & \textbf{VGG11 (sWS)} & \textbf{9.23M} & \textbf{20} & \textbf{97.92\%} & \textbf{208.54} & \textbf{47,133.58} \\
& \textbf{SLTT} & \textbf{VGG11 (sWS)} & \textbf{9.23M} & \textbf{10} & \textbf{96.88\%} & \textbf{120.74} & \textbf{25,468.13} \\
& SpikON (ours) & VGG11 (LTTT) & 9.23M & 20 & 97.57\% & 120.63 & 28,646.23 \\
& \textbf{SpikON (ours)} & \textbf{VGG11 (LTTT+sWCTT)} & \textbf{9.23M} & \textbf{20} & \textbf{97.22\%} & \textbf{156.02} & \textbf{36,603.03} \\
& \textbf{SpikON (ours)} & \textbf{VGG11 (LTTT+sWCTT)} & \textbf{9.23M} & \textbf{10} & \textbf{97.57\%} & \textbf{80.84} & \textbf{18,203.50} \\
\bottomrule
\end{tabular}}
\vspace{-10pt}
\end{table*}

\textbf{Spike similarity.} We compare the spike differences between adjacent timesteps in the forward pass by calculating the cosine similarity~(Fig.~\ref{fig6}). For CIFAR-100, all layers exhibit consistently high cosine similarity~(${>}80\%$), as the dataset is static and each timestep receives identical input. In contrast, the inputs vary dynamically across timesteps for DVS-CIFAR10, leading to lower similarity in certain layers. However, multiple layers~(e.g., layers 2, 4, 6, 7, and 8) still preserve more than $55\%$ similarity across timesteps.

\textbf{CTCR strategy.} Owing to the high temporal similarity of spike activations, we propose the CTCR method to reuse intermediate results across adjacent timesteps~(green circles and arrows in Fig.~\ref{fig5}(a)). Specifically, the convolution result of $s^l_t$ with $w^l$~(denoted as $s^l_t{\ast} w^l$) can be reused for computing $s^l_{t+1}{\ast} w^l$, where only the difference term $(s^l_{t+1}{-}s^l_t){\ast}w^l$ needs to be processed. Since $(s^l_{t+1}{-}s^l_t)$ introduces higher sparsity than $s^l_{t+1}$, the proposed PU~(Fig.~\ref{fig5}(b)) can utilize this property to reduce computation energy. Note that the same reuse mechanism can also be applied to FC layers. The two theorems that support our CTCR strategy are presented below:
\begin{align}
\textit{FC layer: } y^l_t&{=}w^ls^{l-1}_t{=}w^ls^{l-1}_{t-1}{+}w^l(s^{l-1}_t{-}s^{l-1}_{t-1}),\\
\textit{Conv. layer: } y^l_t&{=}w^l{\ast}s^{l-1}_t{=}w^l{\ast}s^{l-1}_{t-1}{+}w^l{\ast}(s^{l-1}_t{-}s^{l-1}_{t-1}),
\end{align}
where $y^l_t$ represents the intermediate result. To verify the sparsity improvement introduced by the CTCR strategy, we conduct the following theoretical analysis.
\begin{align*}
&\textit{Assume the shape of $s_t$, $s_{t+1}$ is (1, D), and the non-zero ratio of $s_t$} \\
&\textit{and $s_{t+1}$ is m and n, respectively. Define $\Delta{=}s_{t+1}{-}s_t$}.\\
&\textit{Similarity: $\cos(s_t, s_{t+1}){=}\frac{\|s_t\|^2{+}\|s_{t+1}\|^2{-}\|s_{t+1}{-}s_t\|^2}{2 \|s_t\| \|s_{t+1}\|}{=}c$.} \\
&\textit{Given $\|s_t\| {=} \sqrt{mD}$ and $\|s_{t+1}\| {=} \sqrt{nD}$, we will have}\\
&\textit{$c{=}\frac{mD {+} nD {-} \|\Delta\|^2}{2D\sqrt{mn}}$, $\|\Delta\|^2 {=} (m {+} n)D {-} 2cD\sqrt{mn}$.}
\end{align*}
Therefore, the non-zero ratio of $\Delta$ is $(m {+} n) {-} 2c\sqrt{mn}$. To guarantee the sparsity improvement of CTCR, it needs to satisfy $\sqrt{m} {\leq} 2c\sqrt{n}$.%~(details are shown in Section~\ref{section5_2}).

\subsection{Customized SIMD-based SNN Core}
To efficiently support neuron operations and ensure the generality of the SpikON accelerator, we design a SIMD-based SNN core with a five-stage pipeline and a lightweight custom ISA. As shown in Fig.~\ref{fig5}, SNN core includes the $1024{\times}32bit$ instruction SRAM, $32{\times}512bit$ register file, executor with 64 FP32 units for parallel computation, controller, etc. Each FP32 unit can execute general arithmetic~(e.g., add, sub, mult, div, sqrt, max, min, etc.) and SNN-specific operations. In particular, \texttt{lif} and \texttt{reset} instructions model LIF neuron dynamics in the forward pass, while \texttt{sg} handles surrogate gradient computation in the backward pass. Moreover, SIMD controller manages pipeline stalls and resolves instruction dependencies to ensure correct and efficient execution. Note that although SNN core executes instructions sequentially without parallel timestep processing as in BTPE, its latency can be effectively hidden since BTPE dominates the overall computation.

\vspace{-1pt}
\section{Experimental Results and Analysis}

\begin{figure*}[tb]
    \centering
    \setlength{\abovecaptionskip}{0pt}
    \includegraphics[width=0.99\linewidth]{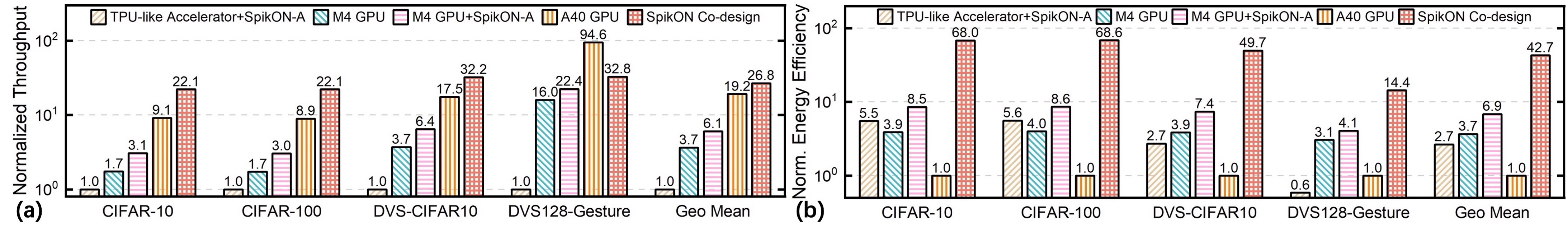}
    \vspace{-1pt}
    \caption{Normalized training throughput~(iteration/s)~(a) and energy efficiency~(iteration/J)~(b) of SpikON co-design over the edge/server GPUs and TPU-like training accelerator under VGG11 SNN model across both static and neuromorphic datasets.} 
    \Description{fig8}
    \label{fig8}
    \vspace{-12pt}
\end{figure*}

\subsection{Experimental Setup}
\textbf{Algorithm setup.} We evaluate the proposed SpikON algorithm on both static and neuromorphic datasets, such as CIFAR-10~\cite{krizhevsky2009learning}, CIFAR-100~\cite{krizhevsky2009learning}, DVS-CIFAR10~\cite{li2017cifar10}, and DVS128-Gesture~\cite{amir2017low}. We use the VGG11 SNN architecture with a lightweight classifier~(64C3-128C3-AP2-256C3-256C3-AP2-512C3-512C3-AP2-512C3-512C3-GAP-FC) for all experiments~\cite{simonyan2014very}. The training epoch and batch size of SNNs are set to 100 and 1, respectively. The leaky constant $\beta$ is configured as 0.09 for our SNN models. The baselines are the prior online SNN learning algorithms, OTTT~\cite{xiao2022online} and SLTT~\cite{meng2023towards}. 

\textbf{Hardware setup and baseline.} We design the SpikON accelerator at RTL using Verilog. To estimate the area and power of core circuits, we synthesize SpikON using Synopsys Design Compiler~\cite{synopsys_dc} at 1.5 GHz under the TSMC 28nm standard cell library. The on-chip SRAMs are generated by the TSMC memory compiler. The energy consumption of off-chip memory HBM2~(5.7pJ/b) is adopted from \cite{chatterjee2017architecting}. In addition, we implement a simulator to evaluate the performance of the SpikON accelerator. SpikON is benchmarked against the edge Apple M4 GPU~(10 cores)~\cite{apple_m4} and server Nvidia A40 GPU~\cite{nvidia_ampere}. We also build a TPU-like training accelerator with 1521 floating-point PEs~(systolic array replacing BTPE module in Fig.~\ref{fig5}) as a baseline and evaluate it under our simulator framework.

\begin{figure}[tb]
    \centering
    \setlength{\abovecaptionskip}{0pt}
    \includegraphics[width=0.95\linewidth]{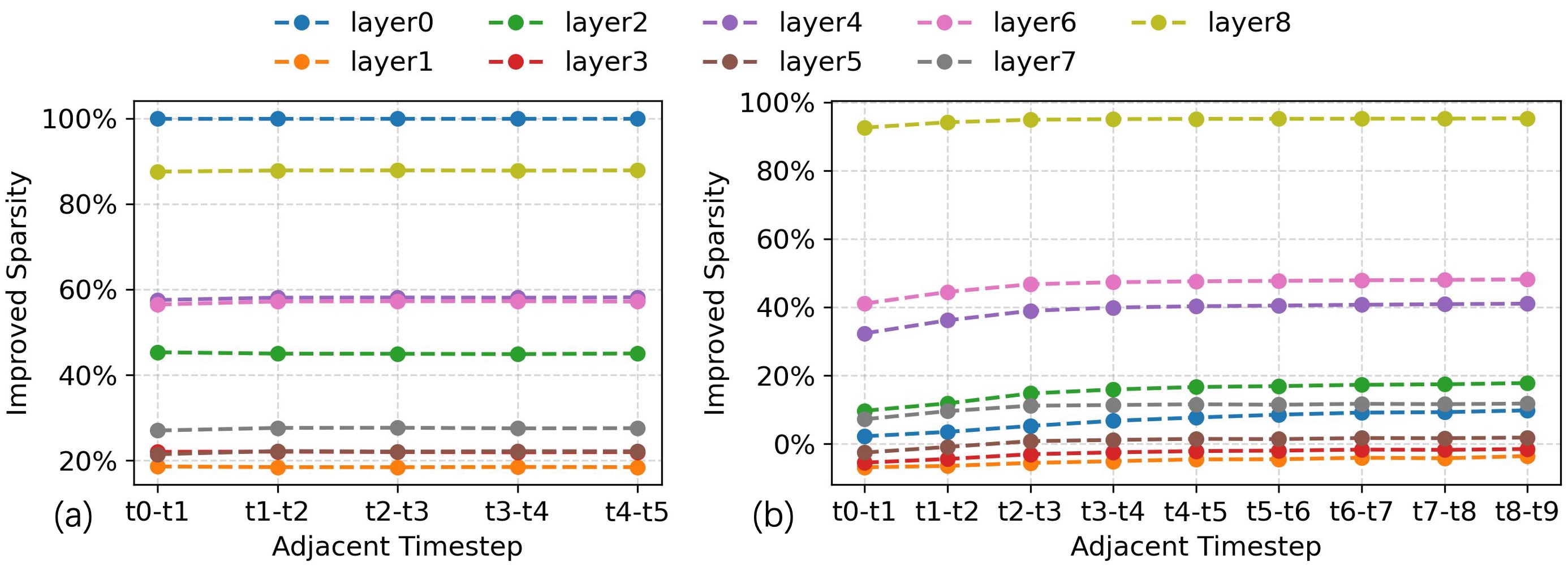}
    %\vspace{-1pt}
    \caption{Sparsity improvement between adjacent timesteps on CIFAR-100~(a) and DVS-CIFAR10~(b) datasets using the proposed CTCR method~(VGG11).}
    \Description{fig7}
    \label{fig7}
    \vspace{-12pt}
\end{figure}

\begin{figure}[tb]
    \centering
    \setlength{\abovecaptionskip}{0pt}
    \includegraphics[width=0.94\linewidth]{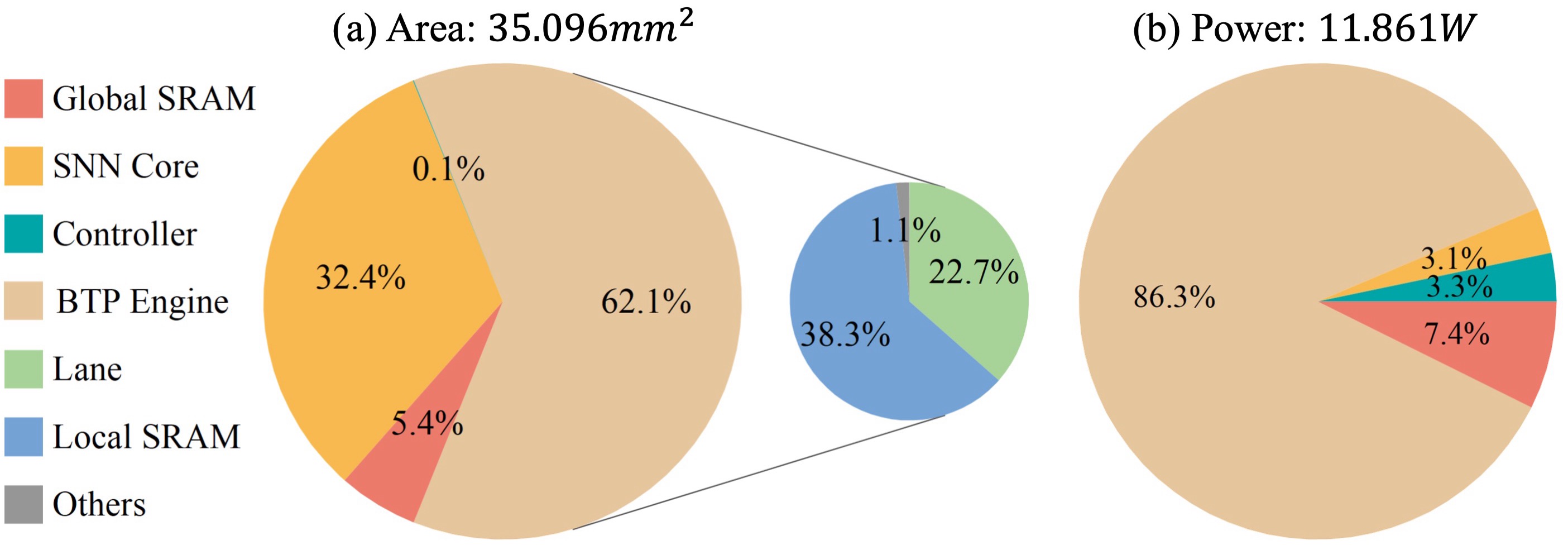}
    %\vspace{-1pt}
    \caption{The area~(a) and power~(b) breakdown of SpikON.}
    \Description{fig9}
    \label{fig9}
    \vspace{-12pt}
\end{figure}

\vspace{-3pt}
\subsection{SpikON Training Algorithm Performance}
\label{section5_2}
\textbf{Comparison with OTTT and SLTT.} \underline{(1) Accuracy.} As shown in Table~\ref{table1}, SpikON~(LTTT-only) exhibits an average accuracy drop of 4.54\% compared to the SLTT~(sWS) baseline across all datasets. However, by further incorporating the sWCTT strategy, SpikON~(LTTT+ sWCTT) even surpasses the SLTT~(sWS), achieving an average accuracy improvement of 0.08\%. \underline{(2) Latency.} We can observe that SLTT~(sWS) achieves a $30.4\%$ reduction on average in training latency over the OTTT~(sWS). Moreover, our SpikON~(LTTT+sWCTT) further reduces the training latency by 32.2\% compared to SLTT~(sWS), without sacrificing accuracy. \underline{(3)} \underline{Energy.} In terms of training energy, SLTT~(sWS) consumes $27.8\%$ less energy on average than OTTT~(sWS). Benefiting from the normalization-free online learning, SpikON~(LTTT+sWCTT) achieves a further $35.0\%$ reduction in energy consumption relative to SLTT~(sWS). 

\textbf{Sparsity improvement achieved by CTCR.} To verify the effectiveness of CTCR, we evaluate the improved sparsity between adjacent timesteps across all layers on CIFAR-100 and DVS-CIFAR10 during training~(Fig.~\ref{fig7}). We can observe that CTCR consistently enhances the sparsity of all layers for the static CIFAR-100. In particular, the improved sparsity of $layer0$ reaches 100\% since it receives the same input at each timestep. Moreover, most layers~(e.g., 0, 2, 3, 6, 7, 8) exhibit increased sparsity under CTCR on the DVS-CIFAR10, with only minor reductions observed in a few layers. Note that the computation overhead of CTCR is negligible, as it only involves subtracting activations between adjacent timesteps.

\subsection{SpikON Accelerator Evaluation}

\textbf{Area and power breakdown.} Fig.~\ref{fig9} presents the area and power breakdown of the SpikON accelerator. The total area and power consumption are $35.096$ $mm^2$ and $11.861$ $W$, respectively. \underline{(1) Area.} The SNN core and BTP engine occupy the majority of the chip area~(32.4\% and 62.1\%), while the controller contributes only 0.1\% of the total area. Within the BTP engine, most of the space is allocated to the 24 BTP-dataflow lanes and 23 local SRAMs, taking up 22.7\% and 38.3\% respectively. \underline{(2) Power.} The BTP engine dominates the overall power, accounting for 83.6\% of the total power. The SNN core and global SRAM consume only 3.1\% and 7.4\%, respectively. 

\begin{figure}[tb]
    \centering
    \setlength{\abovecaptionskip}{0pt}
    \includegraphics[width=\linewidth]{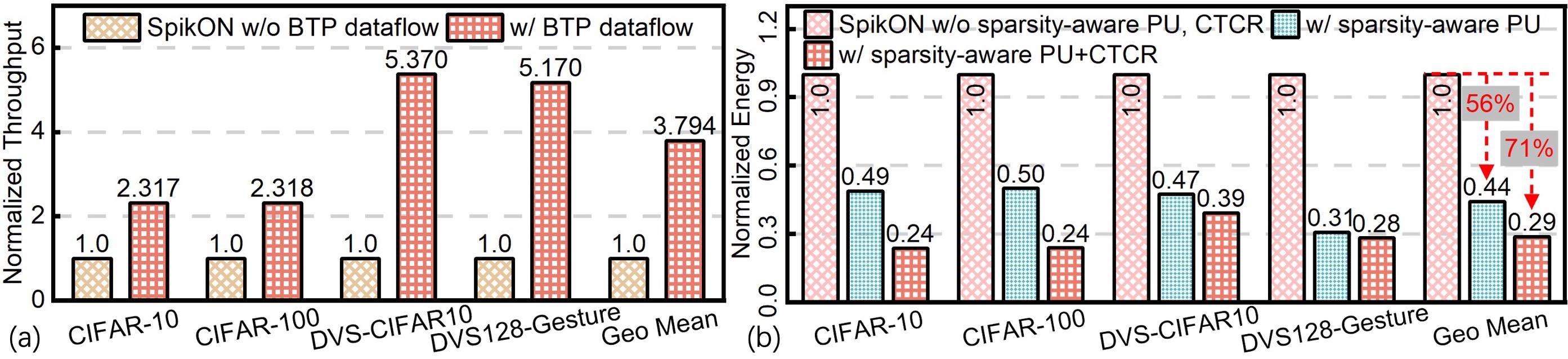}
    \vspace{-10pt}
    \caption{Ablation study of BTP dataflow, sparsity-aware PU, and CTCR on training throughput~(a) and energy~(b).}
    \Description{fig10}
    \label{fig10}
    \vspace{-18pt}
\end{figure}

\textbf{Comparison with edge/server GPUs and TPU-like accelerator.} We use Apple's $powermetrics$ tool~\cite{hubner2025apple} and $nvidia$-$smi$~\cite{nvidia_smi} to measure the power of the M4 and A40 GPUs at runtime, respectively. \underline{(1) SpikON algorithm+M4 GPU.} With the proposed SpikON algorithm, the online SNN training throughput and energy efficiency of the edge M4 GPU are improved by 1.7x and 1.9x, respectively~(Fig.~\ref{fig8}). \underline{(2) SpikON co-design v.s. GPUs.} With both the SpikON algorithm and accelerator, SpikON achieves 7.2x throughput and 11.5x energy efficiency on average over the M4 GPU~(Fig.~\ref{fig8}). Even compared with the server A40 GPU, SpikON still delivers 1.4x and 42.7x improvement in throughput and energy efficiency, respectively. Note that on the DVS128-Gesture dataset, the throughput of SpikON co-design~(32.8x) is lower than A40 GPU~(94.6x), mainly due to the large 128$\times$128 input feature map and the limited parallelism of each lane. However, SpikON still demonstrates a 14.4x advantage in energy efficiency over the A40 GPU. \underline{(3) SpikON co-design v.s. TPU-like} \underline{accelerator.} Due to the lack of an existing SOTA ASIC baseline for online supervised SNN learning, we implement a TPU-like training accelerator and keep all hardware configurations with the same number of PEs for fair comparison. As shown in Fig.~\ref{fig8}, we observe that SpikON co-design achieves 26.8x higher throughput and 15.8x higher energy efficiency on average.

\textbf{Ablation study of SpikON accelerator.} We conduct an ablation study to demonstrate the advantages of the proposed methods~(Fig.~\ref{fig10}). \underline{(1) BTP training dataflow.} As shown in Fig.~\ref{fig10}(a), with the BTP dataflow, we can achieve an average 3.8x speedup in training throughput across all datasets. \underline{(2) Sparsity-aware PU+CTCR.} Fig.~\ref{fig10}(b) presents the energy reduction achieved in the forward pass using the proposed techniques. The sparsity-aware PU utilizes the inherent sparsity of SNNs to reduce energy by 56\% on average. With the additional CTCR strategy, the energy consumption is further reduced to 71\%. Notably, the CTCR achieves greater energy reduction on static datasets than on event-stream datasets, which is aligned with the improved sparsity results in Fig.~\ref{fig7}.

\section{Conclusion}

We propose SpikON, the first algorithm-hardware co-design for efficient and scalable end-to-end online supervised SNN learning. Specifically, SpikON algorithm achieves 32.2\% and 35.0\% reductions in training latency and energy over baseline, without sacrificing accuracy. Moreover, SpikON co-design achieves 7.2x (11.5x) and 26.8x (15.8x) training throughput (energy efficiency) compared with the edge Apple M4 GPU and TPU-like accelerator, respectively.

\bibliographystyle{ACM-Reference-Format}
\bibliography{citations}

\end{document}